\documentclass[runningheads]{llncs}
\usepackage{amsmath}
\usepackage{amsfonts}
\usepackage{amssymb}
\usepackage{graphicx}
\usepackage[ruled]{algorithm2e}
\usepackage{subfigure}

\usepackage{booktabs}

\newcommand{\tabincell}[2]{\begin{tabular}{@{}#1@{}}#2\end{tabular}}
\newcommand{\tablefont}{\fontsize{9pt}{\baselineskip}\selectfont}

\usepackage{color}

\newcommand{\repeatthanks}{\textsuperscript{\thefootnote}}

\begin{document}
\title{I-Nema: A Biological Image Dataset for Nematode Recognition}
%
%
\author{Xuequan Lu\inst{1}\thanks{Joint first author. Q. Xue is the correponding author. This is a preprint.}
\and
Yihao Wang\inst{2}\repeatthanks
\and
Sheldon Fung\inst{1}
\and
Xue Qing\inst{2}
}
\authorrunning{X. Lu et al.}
%
\institute{Deakin University, Australia\\
\and
Nanjing Agricultural University, China
}
\maketitle              
\begin{abstract}
Nematode worms are one of most abundant metazoan groups on the earth, occupying diverse ecological niches. Accurate recognition or identification of nematodes are of great importance for pest control, soil ecology, bio-geography, habitat conservation and against climate changes. Computer vision and image processing have witnessed a few successes in species recognition of nematodes; however, it is still in great demand. In this paper, we identify two main bottlenecks: (1) the lack of a publicly available imaging dataset for diverse species of nematodes (especially the species only found in natural environment) which requires considerable human resources in field work and experts in taxonomy, and (2) the lack of a standard benchmark of state-of-the-art deep learning techniques on this dataset which demands the discipline background in computer science. With these in mind, we propose an image dataset consisting of diverse nematodes (both laboratory cultured and naturally isolated), which, to our knowledge, is the first time in the community. We further set up a species recognition benchmark by employing state-of-the-art deep learning networks on this dataset. We discuss the experimental results, compare the recognition accuracy of different networks, and show the challenges of our dataset. \textit{We make our dataset publicly available at: \url{https://github.com/xuequanlu/I-Nema} } 
\end{abstract}
%
%
%

\section{Introduction}
\label{sec:introduction}

Nematode worms are important, due to the following reasons: (1) parasitic nematodes threaten the health of plants on a global scale, causing at least $80$ billion US dollar's  loss per year \cite{nicol2011current}; (2) interstitial nematodes pervade sediment and soil ecosystems in overwhelming numbers \cite{lambshead1993recent}; and (3) \textit{Caenorhabditis elegans} (\textit{C. elegans}) is a favourable experimental model system \cite{riddle1997developmental}. Accurate recognition or identification of nematodes are of great value for pest control (e.g. choosing a specific nematicide or rotation crop), soil ecology (e.g. evaluation for the soil quality), bio-geography, habitat conservation and climate maintenance \cite{coomans2002present}.  However,  nematodes recognition is challenging, due to high phenotypic plasticity (i.e. high morphological variations within a single species) \cite{coomans2002present,nadler2002species}, vague diagnostic characteristics \cite{derycke2008disentangling,erwin1982tropical}, and frequently encountered juveniles \cite{anderson2000nematode}. More importantly, the manual identification is extremely time-consuming and labor-intensive, especially when dealing with large-scale ecological data. It often requires solid training and well-established experience. As a result, the taxonomy of nematode species appears to be a significant bottleneck for many relevant fields.

\begin{table}[t]\tablefont
    \centering
    \caption{Datasets for Nematode recognition. \#. Samples and \#. Classes denote the numbers of samples and classes, respectively. Our proposed dataset is the first open-access one which involves both naturally isolated and laboratory cultured nematodes, covering diverse morphology and life strategies.  }\label{table:datasetcontrast}
    \begin{tabular}{l c c c}
    \toprule
    Datasets & \tabincell{l}{
    \#. Samples
    } & \tabincell{l}{
    \#. Classes
    } & \tabincell{l}{
    Open to public
    } \\ 
    \midrule
    \tabincell{l}{\cite{liu2010-xray,liu2010-xraymultilinear}} & 50 & 5 & NO
    \\
    \tabincell{l}{\cite{liu2018-cnn}} & 500 & 5 & NO
    \\
    \tabincell{l}{\cite{liu2018-projection}} & 1,000 & 5 & NO
    \\
    \tabincell{l}{\cite{Liu2017-1-imagefusion,liu2017-2-projection}} & 500 & 10 & NO
    \\
    \tabincell{l}{\cite{zhou2006automatic}} & 480,000 frames & 8 & NO
    \\
    \tabincell{l}{\cite{javer2018identification}-single} & 10,476 clips & 365 & NO
    \\
    \tabincell{l}{\cite{javer2018identification}-multi} & 308 clips & 12 & NO
    \\
    \tabincell{l}{\textbf{Our dataset}} & 2,769 & 19 & YES
    \\
    \bottomrule
    \end{tabular}
\end{table}


Although molecular barcoding \cite{floyd2002molecular,blaxter2004promise} has been consolidated as a powerful tool for the species identification and biodiversity investigation, its applications depend on the availability of molecular biology facilities, sequencing instruments, as well as the background knowledge. Alternatively, imaging data are often more accessible and economical for broader users, and have been utilized for nematode research, such as nematode detection/segmentation \cite{silva2003intelligent,nguyen2006improved,ochoa2007contour,rizvandi2008machine,Rizvandi-2,Rizvandi-3,chou2017edge,wang2020celeganser,chen2020cnn}, classification \cite{javer2018identification,zhou2006automatic,liu2010-xray,liu2010-xraymultilinear,liu2017-informationfusion,Liu2017-1-imagefusion,liu2017-2-projection,liu2018-projection,liu2018-cnn}, nematode (eggs) counting \cite{akintayo2018deep,holladay2016high}, etc. It should be noted that the current 
identification or recognition research are either for nematode image stacks \cite{liu2010-xray,liu2010-xraymultilinear,liu2017-informationfusion,Liu2017-1-imagefusion,liu2017-2-projection,liu2018-projection,liu2018-cnn}, or predominantly designed for model organism \textit{ C. elegans} \cite{zhou2006automatic,javer2018identification}. However, the involved issues for identification tasks are: (1) very few classes involved in the image stacks, e.g. $5$ classes in Table \ref{table:datasetcontrast}; (2) low diversity coverage, i.e. species limited to the laboratory cultured \textit{C. elegans} only (Table \ref{table:datasetcontrast}: \cite{zhou2006automatic,javer2018identification}). 
Importantly, to our knowledge, all relevant imaging data were used in  their own research, and none of them are publicly available. In summary, an image dataset covering diverse kinds of nematodes and a standard benchmark using state-of-the-art deep learning techniques is still missing! 

Provided the above analysis, we are motivated to create an image dataset for nematodes and complete a benchmark for state-of-the-art deep learning methods. 
Despite that laboratory cultured species are much easier to acquire, for example, \textit{C. elegans} complete a reproductive life cycle in $3.5$ days at $20^\circ$ and a huge population can be reproduced, our samples are mostly collected from the natural environment (many are not cultivable in laboratory). As a result, the data collection is more time-consuming and labor-intensive (e.g. sampling in different ecosystems, manual nematode picking for target species in a mixed population, etc) than using pure laboratory culture. 
In particular, we first collect soil from a wide range of natural environments, including temperate broadleaf, mixed forest, crop field, and tundra. Then the nematodes are extracted and further placed in glass slides. With the aid of the microscope system, manual identification is performed to determine whether a nematode is needed or not for further image capturing. If it is needed, we continue image capturing using the microscope imaging system (a camera, a microscope and a software), and assign the manually identified species label to the involved images. It is discarded otherwise. We also take images for $2$ laboratory cultured species.
A single nematode worm per image is constrained during image capture. 
The dataset has a total number of $2,769$ images and $19$ different species ($17$ species from natural environment and $2$ laboratory species).  
To the best of our knowledge, \textit{this proposed dataset is the first open-access image dataset including diverse nematodes species and life strategies (plant-parasitic, fungi feeder, bacteria feeder, omnivores, predator)}. 

In addition to the dataset, we further employ the state-of-the-art deep learning networks on the species recognition of nematodes. 
We analyze the results with regard to different respects: (1) different deep learning networks, (2) pre-training as initialization versus training from scratch. We also conduct experiments and analysis on augmentation and discuss the supported research by our work. 
The \textit{contributions} of this work are summarized as follows.
\begin{itemize}
    \item We propose an image dataset for diverse species of nematodes. It is, to our knowledge, the first open-access biological image dataset for diverse nematodes, representing different evolutionary lineages and life strategies (isolated both from natural environment and laboratory). 
    
    \item We conduct a benchmark by adjusting and training the state-of-the-art deep learning networks on our dataset. We compare and analyze their performance. We also discuss the supported research by our work. 
\end{itemize}



\section{Related Work}
\label{sec:relatedwork}

In this section, we review the research which are most relevant to this work. We first look back on the nematode detection/segmentation, and then review species recognition of nematodes. We finally summarize some state-of-the-art CNNs.

\subsection{Nematode Detection/Segmentation}
\label{related-detection}
An early nematode detection paper presented a computational system which supports the detection of nematodes in digital images \cite{silva2003intelligent}. Nguyen et al. proposed an improved watershed segmentation algorithm using water diffusion and local shape priors, which has been applied to the segmentation of nematode worms \cite{nguyen2006improved}. Ochoa et al. introduced an approach to extract a considerable number of individuals even in cluttered regions in population images \cite{ochoa2007contour}. Later, novel methods based on skeleton analysis were designed for the detection of individual nematode worms in population images in presence of worm overlapping \cite{rizvandi2008machine,Rizvandi-2,Rizvandi-3}. Recently, Chou et al. put forwarded an efficient CNN-based regression network for accurate edge detection \cite{chou2017edge}. Chen et al. proposed a framework for detecting worm-shaped objects in microscopic images using convolutional neural networks (CNNs) \cite{chen2020cnn}. The authors used curved lines as annotation rather than bounding boxes. Wang et al. introduced a pipeline for automated analysis of C. elegans imagery \cite{wang2020celeganser}, which detects, segments the worm, and predicts the age of individual worms with high accuracy.

\begin{figure*}[thbp]
\centering
\begin{minipage}[b]{1.0\linewidth}
{\label{}\includegraphics[width=1\linewidth]{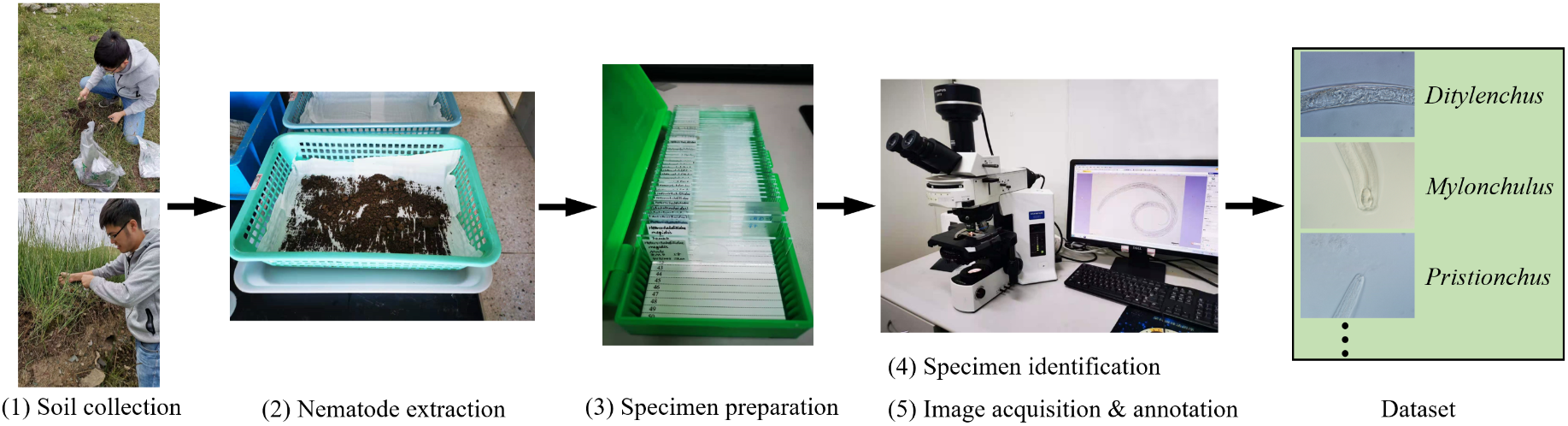}}
\end{minipage}
\caption{ The pipeline for creating our dataset. Soil samples are first collected (soil collection), and nematodes are then extracted from soil (extraction). The nematodes are further picked up and placed in a glass slide (specimen preparation). After manual identification (identification), we determine which specimens will do image capturing and annotation (image acquisition and annotation).   }
\label{fig:overview}
\end{figure*}

\subsection{Nematode Recognition}
\label{related-classification}
A major proportion of nematode classification research are conducted on image stacks, with each stack (or each species) consisting of multiple focal planes of the specimen \cite{liu2010-xray,liu2010-xraymultilinear,liu2017-informationfusion,liu2018-projection,liu2018-cnn,Liu2017-1-imagefusion,liu2017-2-projection}. Various methods have been developed to handle the classification based on image stacks, for example, information fusion based approaches \cite{liu2017-informationfusion,Liu2017-1-imagefusion}, 3D X-Ray Transform based method \cite{liu2010-xray}, 3D X-Ray Transform based multilinear analysis method \cite{liu2010-xraymultilinear}, projection-based methods \cite{liu2017-2-projection,liu2018-projection}, deep convolutional neural network (CNN) image fusion based multilinear approach \cite{liu2018-cnn}. Zhou et al. proposed an identification method for mutant types and other types, based on locomotion patterns \cite{zhou2006automatic}. Javer et al. presented a fully convolutional neural network to discern genetically diverse strains of C. elegans, based only on their recorded spontaneous activity, and explored how its performance changes as different embeddings are used as input \cite{javer2018identification}.

\textbf{Datasets.} \cite{liu2010-xray,liu2010-xraymultilinear}
used $5$ species and $10$ samples per species for classification. \cite{liu2018-cnn} adopted $500$ samples from $5$ classes, and \cite{liu2018-projection} utilized $1,000$ samples from $5$ species. \cite{Liu2017-1-imagefusion,liu2017-2-projection} used $500$ samples of $10$ categories.  \cite{zhou2006automatic} used a total number of $480,000$ frames from the wild type and $7$ other mutant types of C. elegans. \cite{javer2018identification} involves two datasets of C. elegans: single-worm dataset and multi-worm dataset. The former includes $10,476$ video clips of individual worms divided between $365$ different classes. The latter contains a total number of $308$ video clips from $12$ strains. We summarize the dataset information in Table \ref{table:datasetcontrast}. \textit{It is not surprising to collect a great number of samples for C. elegans, since C. elegans is a common organism with rapid reproduction in laboratory. Also, the data are typically used in their own research and are not  released to the public. }

\subsection{Deep CNNs}
\label{related-cnns}
Nowadays, deep convolutional neural networks (CNNs) are powerful tools for data-driven research. State-of-the-art CNNs are AlexNet \cite{alexnet}, VGG \cite{vgg}, ResNet \cite{resnet}, Inception \cite{inception}, Xception \cite{xception}, etc. They are well defined and built on image grids, and capable of learning discriminative features from input images. In this work, we adjust AlexNet, VGG-16, VGG-19, ResNet-34, ResNet-50, ResNet-101 and train them on the train set, and test them over the test set of our dataset, in terms of the task of nematode recognition.

\section{Our Dataset: I-Nema}
\label{sec:dataset}

\subsection{Overview}
\label{sec:overview}
In this section, we explain how we create our dataset, referring to as ``I-Nema''. At first, we conduct a country-wide soil sample collection in order to gather specimens representing different lineages in the tree of Nematoda \cite{holterman2006phylum}. The individual nematodes are extracted from soil, fixed, and subsequently processed into glass slide specimens with $10\sim20$ nematode worms per glass slide. The species is manually identified and the undesirable taxa are excluded. Images are then captured for each selected specimen using the camera equipped with a microscope. 
The identified species label is simply annotated to the involved images. We finally perform some pre-processing for the initially collected images. Figure \ref{fig:overview} shows the pipeline for creating I-Nema.

\subsection{Sample Collection}
\label{sec:samplecollection}
Sample collection  contains soil collection, nematode extraction and specimen preparation.
Soil samples are collected from various ecosystems in Mainland China. Aside from collecting samples in the natural environment, we also use the nematode cultures maintained in laboratory (two species: \textit{Pratylenchus} sp. and \textit{Acrobeloides} sp.). For soil samples, nematodes are extracted with a Baermann tray after an incubation in room temperature for 24 hours. The nematode extraction are concentrated using a 500 mesh sieve (25µm opening). For the laboratory maintained species, nematodes are directly washed from either carrot disc (\textit{Pratylenchus} sp.) or cultural middle (\textit{Acrobeloides} sp.).  After removing water, nematodes are fixed with $4\%$ formalin and gradually transferred to anhydrous glycerin, following the protocol \cite{sohlenius1987vertical} which is modified based on \cite{seinhorst1962killing}. The glycerin preserved nematodes are manually picked up and placed in a glass slide. Each permanent slide contains about $10\sim20$ individual worms. 

\subsection{Specimen Recognition}
\label{sec:specimenid}
The species is manually identified before capturing images. This is to collect diverse species of nematodes, rather than simply increasing the number of images in a few species. Nematodes are identified based on the morphological characters (e.g. the presence of teeth or stylet in buccal cavity, the shape of pharynx and tail, the type of reproductive system, etc) and morphometric measurements (e.g. the length of stylet, body width and length, tail length, etc). These characters can be straightforwardly observed through microscopy and/or measured in a professional software connected to the camera equipped with the microscope. The acquired information is subject to the available taxonomic key (e.g. \cite{bongers1988nematoden,andrassy2005free}) and is further confirmed with the original descriptions. The recovered taxon will be considered as undesirable and excluded if it is rare in population (difficult to acquire sufficient specimens) or evolutionary repetitive to a captured species. If the taxon is abundant in number and represent a different evolutionary lineage, we continue the following image acquisition step. We selected and identified a total number of $2$ laboratory cultured species (\textit{Pratylenchus} sp. and \textit{Acrobeloides }sp.) and $17$ naturally collected species (see Table \ref{table:datasetstatistics}), covering $16$ families of common soil species and all nematode trophic guilds. Figure \ref{fig:overview} shows the microscope setup for the manual specimen identification.

\begin{figure}[htbp]
\centering
\begin{minipage}[b]{0.32\linewidth}
\subfigure[]{\label{}\includegraphics[width=1\linewidth]{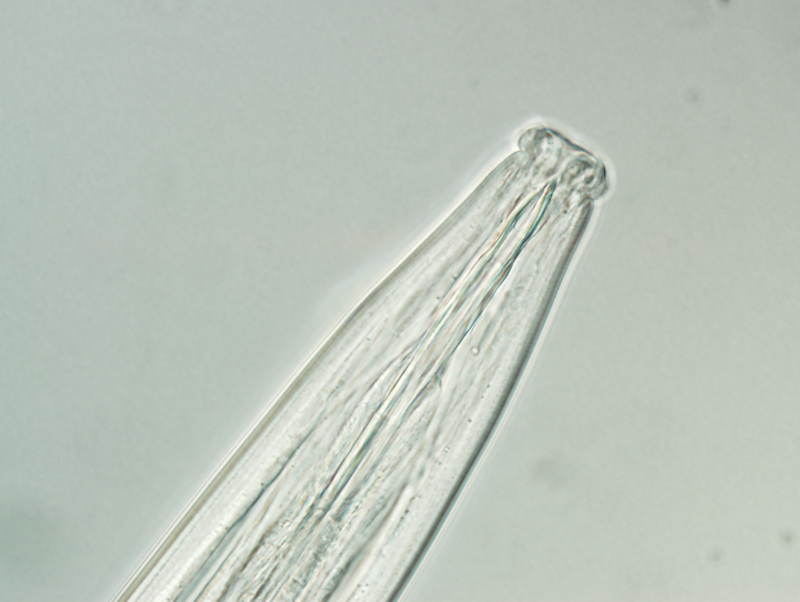}}
\end{minipage}
\begin{minipage}[b]{0.32\linewidth}
\subfigure[]{\label{}\includegraphics[width=1\linewidth]{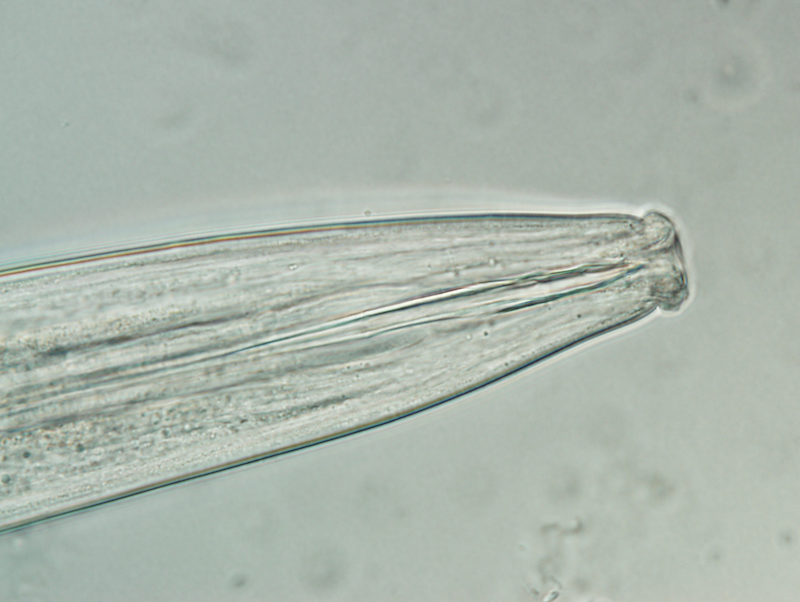}}
\end{minipage}
\begin{minipage}[b]{0.32\linewidth}
\subfigure[]{\label{}\includegraphics[width=1\linewidth]{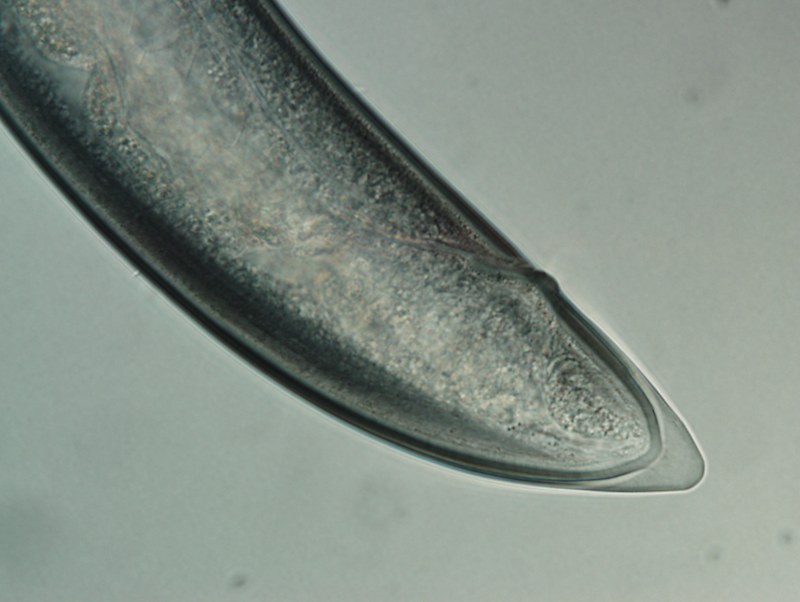}}
\end{minipage} \\
\begin{minipage}[b]{0.32\linewidth}
\subfigure[
]{\label{}\includegraphics[width=1\linewidth]{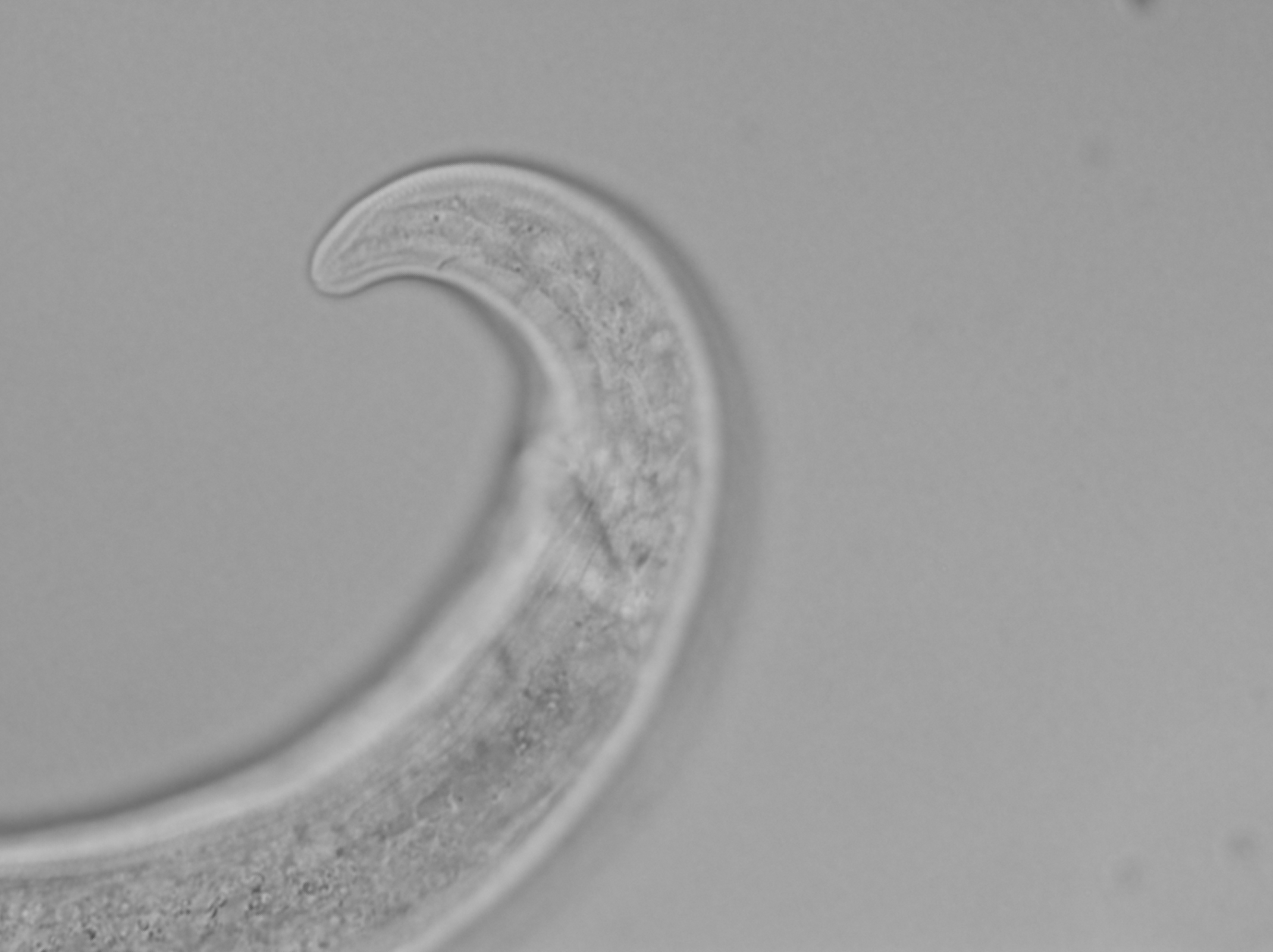}}
\end{minipage}
\begin{minipage}[b]{0.32\linewidth}
\subfigure[
]{\label{}\includegraphics[width=1\linewidth]{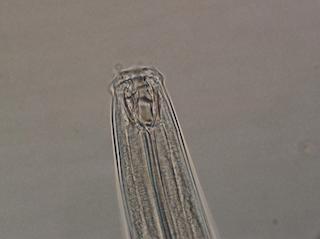}}
\end{minipage}
\begin{minipage}[b]{0.32\linewidth}
\subfigure[
]{\label{}\includegraphics[width=1\linewidth]{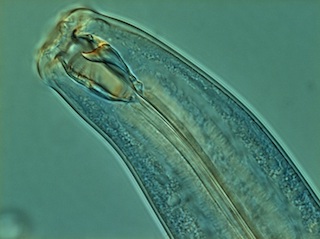}}
\end{minipage} \\
\begin{minipage}[b]{0.32\linewidth}
\subfigure[
]{\label{}\includegraphics[width=1\linewidth]{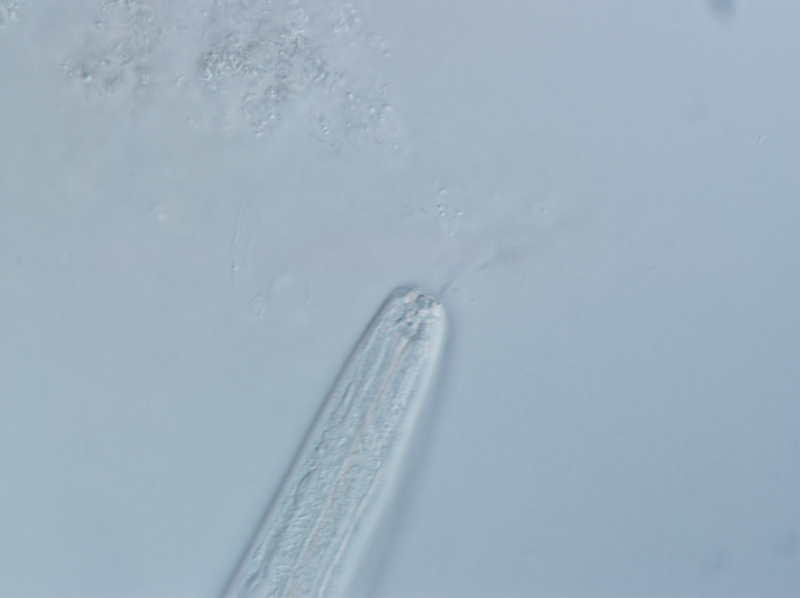}}
\end{minipage}
\begin{minipage}[b]{0.32\linewidth}
\subfigure[
]{\label{}\includegraphics[width=1\linewidth]{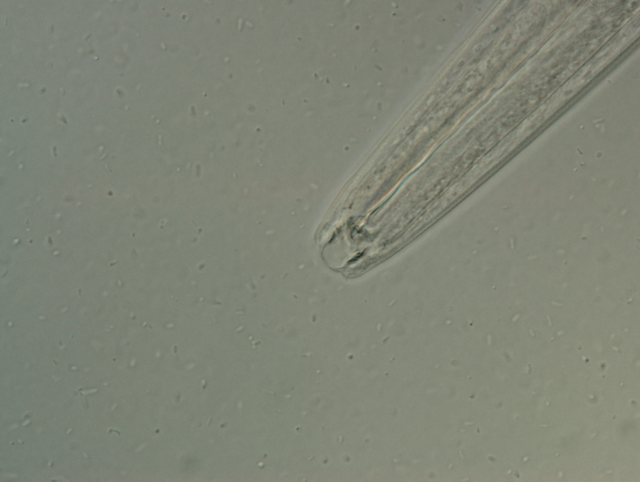}}
\end{minipage}
\begin{minipage}[b]{0.32\linewidth}
\subfigure[
]{\label{}\includegraphics[width=1\linewidth]{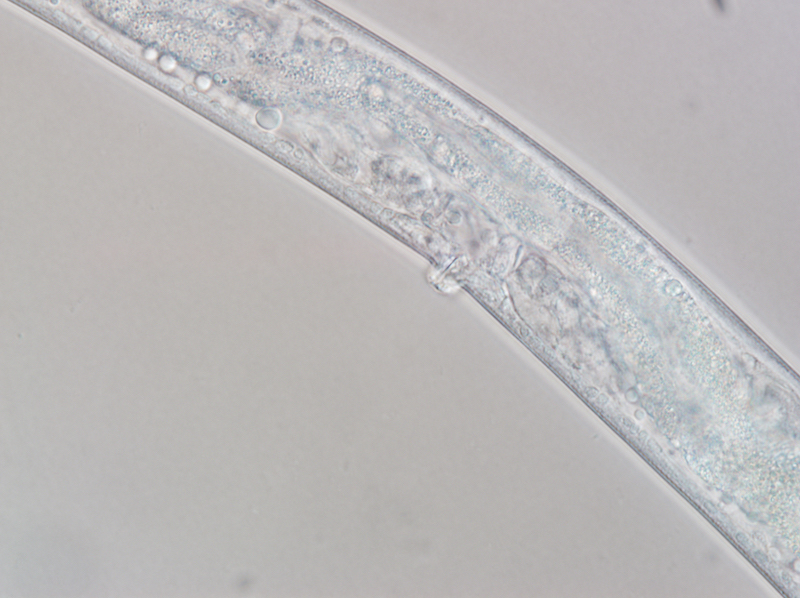}}
\end{minipage}
\caption{ The example of nematode images used in this study. (a-c): \textit{Aporcelaimus} sp.; (d-f): \textit{Mylonchulus} sp.; (g-i): \textit{Pristionchus} sp.; (a), (b), (e), (f), (g), (h): head region; (c,d): tail region; (i): middle body. }
\label{fig:sampleimages}
\end{figure}

\subsection{Image Acquisition and Annotation}
We capture images after the step of identifying specimens. The glass slides are examined and photographed using an Olympus BX51 DIC Microscope (Olympus Optical, produced in Tokyo, Japan), equipped with an camera. The system (see Figure \ref{fig:overview}) has also been utilized for specimen identification in Section \ref{sec:specimenid}. The procedure for image acquisition are as follows. The nematode in specimen is moved to the view center of the microscope, to facilitate further image alignment. 
Head regions are given a higher priority, as they involve more informative taxonomic characters. Tail and middle body regions are also photographed, since these regions are also informative for species recognition and easy to locate.  For the \textit{Mylonchulus} sp., \textit{Miconchus} sp., \textit{Pristionchus} sp. and \textit{Acrobeles }sp., $5\sim10$ images are taken at different image planes to ensure each of the taxonomic characters are properly captured. These species have contrasting morphological structures at different layers.  For other species, $3\sim5$ images are taken from the lowest, middle and highest image planes which allow large variations to be captured. Both juveniles and adult worms are included for capturing images.

\begin{figure}[htbp]
\centering
\begin{minipage}[b]{0.9\linewidth}
{\label{}\includegraphics[width=1\linewidth]{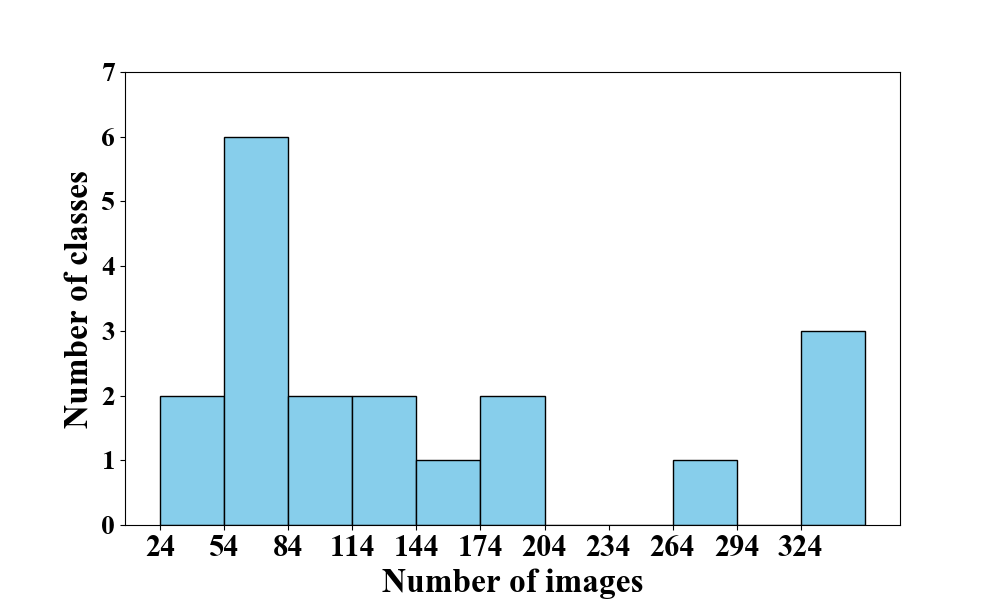}}
\end{minipage}
\caption{ Histogram for our proposed dataset (number of images versus number of classes). }
\label{fig:imagehistogram}
\end{figure}

We eventually acquire a total number of $2,769$ images, ranging from $24$ to $347$ images per species. As for image annotation, since the species of a specimen has been identified in the last step, the species label can be simply allocated to the acquired images of corresponding nematodes. Table \ref{table:datasetstatistics} shows some statistics of our proposed image dataset (I-Nema). Figure \ref{fig:sampleimages} shows some nematode images from I-Nema.
As shown in the histogram (Figure \ref{fig:imagehistogram}), 
the data distribution is diverse. Species with less than $114$ images account for more than half of the whole dataset, while over $324$ images are included in three species. Two species have less than $54$ images. Species between $84$ images and $294$ images have few variations (except blanks), in terms of the number of classes. 

\begin{table}[thbp]
    \centering
    \caption{ Numbers of image samples in each species in our dataset. ``sp.'' represents ``species''. ``\#. Samples'' denotes the number of samples. }\label{table:datasetstatistics}
    \begin{tabular}{l c}
    \toprule
    Species & \#. Samples
    \\ 
    \midrule
    \tabincell{l}{\textit{Acrobeles} sp.} &  71
    \\
    \tabincell{l}{\textit{Acrobeloides} sp.} &  184
    \\
    \tabincell{l}{\textit{Amplimerlinius} sp.} &  24
    \\
    \tabincell{l}{\textit{Aphelenchoides} sp.} &  347
    \\
    \tabincell{l}{\textit{Aporcelaimus} sp.} &  128
    \\
    \tabincell{l}{\textit{Axonchium} sp.} &  170
    \\
    \tabincell{l}{\textit{Discolimus} sp.} &  64
    \\
    \tabincell{l}{\textit{Ditylenchus} sp.} &  330
    \\
    \tabincell{l}{\textit{Dorylaimus} sp.} &  38
    \\
    \tabincell{l}{{\textit{Eudorylaimus}} sp.} &  86
    \\
    \tabincell{l}{\textit{Helicotylenchus} sp.} &  77
    \\
    \tabincell{l}{\textit{Mesodorylaimus} sp.} &  96
    \\
    \tabincell{l}{\textit{Miconchus} sp.} &  57
    \\
    \tabincell{l}{\textit{Mylonchulus} sp.} &  139
    \\
    \tabincell{l}{\textit{Panagrolaimus} sp.} &  326
    \\
    \tabincell{l}{\textit{Pratylenchus} sp.} &  286
    \\
    \tabincell{l}{\textit{Pristionchus} sp.} &  196
    \\
    \tabincell{l}{\textit{Rhbiditis} sp.} &  81
    \\
    \tabincell{l}{\textit{Xenocriconema} sp.} &  69
    \\
    \tabincell{l}{\textbf{Total}} &  2,769
    \\
    \bottomrule
    \end{tabular}
\end{table}

\subsection{Image Pre-processing}
\label{sec:preprocessing}
\begin{figure}[htbp]
\centering
\begin{minipage}[b]{1.0\linewidth}
{\label{}\includegraphics[width=1\linewidth]{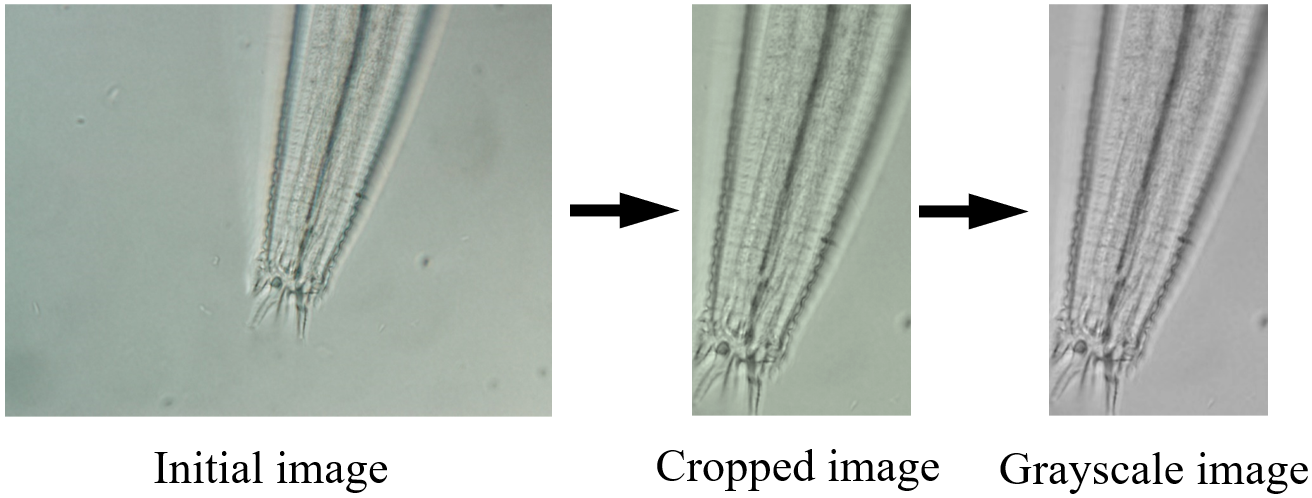}}
\end{minipage}
\caption{ Preprocessing for the initially captured images. We fist crop the less informative regions and then process it into a grayscale image.  }
\label{fig:preprocessing}
\end{figure}

The captured images suffer from several issues, for example, 
nematode worms existing in different regions on images, various background color. 
To alleviate these issues, we first 
crop the images based on the detected edge information. Specifically, we use Canny edge detector \cite{canny1986computational} which is sufficient to detect edges in each image (i.e. worm area), and calculate the min/max pixels for the detected edges. We crop each image according to the corresponding min/max pixels. Note that this is not a worm segmentation, as shown in Figure \ref{fig:preprocessing}. We then convert the cropped images to grayscale images.  
Figure \ref{fig:preprocessing} illustrates the procedure for the image pre-processing step.

\section{Benchmark}
\label{sec:benchmark}

\subsection{Experimental Setting}

\textbf{Data.} We split our proposed dataset into the train set and test set, with an approximate ratio of $4:1$.  
With respect to data augmentation, we utilize online random flip (vertical and horizontal) and Gaussian blur during training. In order to fit the input size of the networks, all images are resized to $224 \times 224$.

\textbf{CNNs.} We conduct two types of experiments using 6 state-of-the-art CNNs which are AlexNet, VGG-16, VGG-19, ResNet-34, ResNet-50 and ResNet-101. Two types are pre-training initialization and training from scratch, respectively. All CNNs are imported from Pytorch \cite{pytorch}. In the case of pre-training initialization, since all networks are pre-trained on the ImageNet \cite{imagenet_cvpr09} which includes $1,000$ classes, we adjust their final output layers to $19$ accordingly, i.e. the number of species in our dataset. 
For simplicity  and consistency, we set the same hyperparameters for all CNNs, including a learning rate of $1.0\times10^{-3}$, a weight decay of $10^{-4}$, 
a batch size of $32$, a cross-entropy loss function and SGD as the optimizer. We assign $100$ epochs for all CNNs, in terms of the training with pre-training initialization.  We assign $150$ epochs for all CNNs, in the case of training from scratch. In both cases, we select the trained models with the best performance for fair comparisons.
We use a desktop workstation with one NVIDIA TITAN V and one NVIDIA QUADRO RTX 6000 for training. 

\textbf{Evaluation metric.}
To evaluate the classification or recognition outcomes, we employ two common metrics which are \textit{Mean Class Accuracy} and \textit{Overall Accuracy}. We use Mean Cls. Acc. and Overall Acc. to represent them, respectively, in the tables. In particular, Mean Class Accuracy is to evaluate the average accuracy of all classes, while Overall Accuracy is for evaluating the average accuracy over all sample images. They are defined as
\begin{equation}\label{eq:metric}
\begin{aligned}
acc_{class} = \frac{1}{c}\sum_{i=1}^c \frac{1}{n_i} \sum_{j=1}^{n_i} a_j^i,\\
acc_{overall} = \frac{\sum_{i=1}^c{\sum_{j=1}^{n_i}} a_j^i } {\sum_{i=1}^c n_i},
\end{aligned}
\end{equation}
where $acc_{class}$ denotes Mean Class Accuracy, and $acc_{overall}$ indicates Overall Accuracy. $c$ is the number of classes, i.e. $19$ in this work. $n_i$ is the number of images in the $i$-th class, and $a_j^i$ is the accuracy for the $j$-th image in the $i$-th class. It should be noted that the train set and test set of our dataset are separate for the metric calculation, and we report the evaluation results of these two metrics for the test set.




\begin{table*}[thbp]
    \centering
    \caption{ Classification or recognition results of different CNNs on our dataset. Numbers before $/$ denote the test results of using pre-trained models (on ImageNet) for training CNNs on our proposed dataset. Numbers after $/$ indicate the test results by training CNNs over our dataset from scratch. All numbers are with $\%$.  We can see that the ResNet series are usually better than other CNNs over our dataset. ``sp.'' represents ``species''. ``Mean Cls. Acc'' and ``Overall Acc.'' denote Mean Class Accuracy and Overall Accuracy, respectively. }\label{table:classificationaccuracy}
    \begin{tabular}{l c c c c c c}
    \toprule
    Species & \tabincell{l}{AlexNet} & \tabincell{l}{VGG16} & \tabincell{l}{VGG19} & \tabincell{l}{ResNet34} & \tabincell{l}{ResNet50} & \tabincell{l}{ResNet101}
    \\ 
    \midrule
\tabincell{l}{\textit{Acrobeles} sp.} & 28.6/0.0 & 50.0/21.4 & 57.1/28.6 & 64.3/28.6 & 64.3/7.1 & 71.4/21.4 
\\
\tabincell{l}{\textit{Acrobeloides} sp.} & 13.9/36.1 & 33.3/16.7 & 58.3/16.7 & 66.7/30.6 & 77.8/27.8 & 80.6/8.3 
\\
\tabincell{l}{\textit{Amplimerlinius} sp.} & 0.0/0.0 & 0.0/0.0 & 0.0/0.0 & 20.0/0.0 & 20.0/0.0 & 20.0/0.0 
\\
\tabincell{l}{\textit{Aphelenchoides} sp.} & 49.3/26.1 & 84.1/37.7 & 92.8/36.2 & 87.0/30.4 & 85.5/65.2 & 91.3/55.1 
\\
\tabincell{l}{\textit{Aporcelaimus} sp.} & 32.0/16.0 & 16.0/32.0 & 8.0/44.0 & 72.0/4.0 & 76.0/8.0 & 36.0/0.0 
\\
\tabincell{l}{\textit{Axonchium} sp.} & 82.4/50.0 & 82.4/52.9 & 79.4/64.7 & 82.4/50.0 & 85.3/47.1 & 91.2/61.8 
\\
\tabincell{l}{\textit{Discolimus} sp.} & 0.0/0.0 & 0.0/0.0 & 0.0/0.0 & 0.0/0.0 & 7.7/0.0 & 0.0/0.0 
\\
\tabincell{l}{\textit{Ditylenchus} sp.} & 73.5/39.7 & 79.4/77.9 & 88.2/69.1 & 72.1/67.7 & 88.2/64.7 & 85.3/57.4 
\\
\tabincell{l}{\textit{Dorylaimus} sp.} & 14.3/0.0 & 28.6/14.3 & 42.9/14.3 & 57.1/14.3 & 42.9/0.0 & 42.9/0.0 
\\
\tabincell{l}{\textit{Eudorylaimus} sp.} & 35.3/0.0 & 82.3/17.7 & 58.8/11.8 & 64.7/23.5 & 64.7/5.9 & 58.8/0.0 
\\
\tabincell{l}{\textit{Helicotylenchus} sp.} & 40.0/0.0 & 53.3/20.0 & 93.3/6.7 & 80.0/26.7 & 93.3/20.0 & 93.3/20.0 
\\
\tabincell{l}{\textit{Mesodorylaimus} sp.} & 10.5/0.0 & 15.8/0.0 & 21.1/0.0 & 47.4/0.0 & 57.9/0.0 & 26.3/0.0 
\\
\tabincell{l}{\textit{Miconchus} sp.} & 45.5/0.0 & 90.9/45.5 & 54.6/18.2 & 90.9/27.3 & 90.9/9.1 & 100.0/18.2 
\\
\tabincell{l}{\textit{Mylonchulus} sp.} & 62.1/0.0 & 89.7/13.8 & 75.9/6.9 & 93.1/0.0 & 96.6/0.0 & 96.6/0.0 
\\
\tabincell{l}{\textit{Panagrolaimus} sp.} & 38.5/23.1 & 61.5/16.9 & 73.9/12.3 & 76.9/18.5 & 73.9/16.9 & 83.1/12.3 
\\
\tabincell{l}{\textit{Pratylenchus} sp.} & 80.0/65.0 & 90.0/63.3 & 88.3/65.0 & 98.3/60.0 & 96.7/60.0 & 98.3/61.7 
\\
\tabincell{l}{\textit{Pristionchus} sp.} & 69.2/7.7 & 74.4/71.8 & 64.1/66.7 & 79.5/59.0 & 69.2/61.5 & 71.8/59.0 
\\
\tabincell{l}{\textit{Rhbiditis} sp.} & 60.0/0.0 & 66.7/20.0 & 53.3/26.7 & 46.7/6.7 & 46.7/46.7 & 93.3/26.7 
\\
\tabincell{l}{\textit{Xenocriconema} sp.} & 36.4/0.0 & 100.0/27.3 & 72.7/27.3 & 100.0/0.0 & 100.0/0.0 & 81.8/0.0 
\\
\tabincell{l}{\textbf{Mean Cls. Acc.}} & 40.6/13.9 & 57.8/28.9 & 57.0/27.1 & 68.4/23.5 & 70.4/23.2 & 69.6/21.1
\\
\tabincell{l}{\textbf{Overall Acc.}} & 50.7/24.6 & 67.0/38.6 & 69.4/36.8 & 76.1/33.3 & 78.6/36.4 & 79.0/32.8
\\
    \bottomrule
    \end{tabular}
\end{table*}

\subsection{Species Recognition}

We perform species identification (or recognition) of nematodes using the six above CNNs on our proposed dataset. The classification results are reported in Table \ref{table:classificationaccuracy}. The results consist of the accuracy for  recognizing each species of nematodes, the mean class accuracy as well as the overall accuracy. Note that the numbers before $/$ are the results by fine-tuning the pre-trained model on our dataset. The numbers after $/$ are obtained through training each CNN on our dataset from scratch. 
It is clear that training from scratch produces inferior test results to those by the fine-tuning with pre-training. On the one hand, it follows the general observation, that is, a good weight initialization could help improve the stability of training and the training outcomes. On the other hand, it also reveals the challenges of our dataset for these state-of-the-art CNNs, which is evidenced by the poor classification accuracy ($<40.0\%$) by training from scratch as well as the limited recognition accuracy ($<80.0\%$) by fine-tuning. 

\subsubsection{Pre-training initialization}

It can be seen from Table \ref{table:classificationaccuracy} that the ResNet series often outperform other CNNs in the case of pre-training initialization. This is probably because the residual concept makes the training of deep networks easier than other CNNs. The highest accuracy is $79.0\%$, achieved by fine-tuning ResNet101. By contrast, AlexNet obtains the lowest recognition accuracy which is $50.7\%$. VGG16 and VGG19 achieve $67.0\%$ and $69.4\%$, respectively. 

With respect to species, we see that the highest test accuracy is $100.0\%$, reported in \textit{Xenocriconema} sp. and \textit{Miconchus} sp., achieved by ResNet family and VGG16. By contrast, nearly all CNNs fail to identify \textit{Discolimus} sp. correctly. This is probably because the images of this species have few discriminative features and CNNs mistake them for other species. 

In general, the species with more images are more likely to be correctly recognized. As shown in Table \ref{table:classificationaccuracy}, the classifiers tend to perform well on the species where sufficient data is provided. Although the number of samples is of importance to the test accuracy, CNNs may also be able to recognize species with less training data. For instance, the number of samples for two different species, \textit{Panagrolaimus} sp. and \textit{Miconchus} sp., are $326$ and $57$ respectively, but most CNNs produced higher test accuracy in identifying \textit{Miconchus}. This may be attributed to high image quality and discriminative features or patterns on the images, which hence provide useful information for CNNs to learn.

\subsubsection{Training from scratch}
Regarding training from scratch, all CNNs lead to poor recognition outcomes which are less than $40.0\%$. Specifically, AlexNet attains a lowest recognition accuracy which is $24.6\%$. Other CNNs are a bit better than AlexNet, achieving over $30.0\%$ accuracy. 

We further noticed that only a few species learned effective features, for example, the species of \textit{Pristionchus}, \textit{Pratylenchus}, \textit{Ditylenchus} and \textit{Axonchium}. We deduce that this is due to a large number of images in these species. By contrast, other species have fewer samples and/or less discriminative features/patterns. We observed the recognition accuracy of some species are even $0\%$. We suspect that our dataset involves challenging patterns or features (e.g. twisted worms, random poses of worms, disturbance information etc) for learning, and a random weight initialization would be too arbitrary to learn these patterns or features. Thanks to the pre-training initialization, we achieve much better recognition outcomes than training from scratch.



\begin{table}[thbp]
    \centering
    \caption{ Comparisons for different augmentation strategies. 
    ``Mean Cls. Acc'' and ``Overall Acc.'' denote Mean Class Accuracy and Overall Accuracy, respectively.  
    }\label{table:augdistricmp}
    \begin{tabular}{l c c c}
    \toprule
    Accuracy (\%) & None &
    \tabincell{l}{Flip} & \tabincell{l}{Flip \& blur} 
    \\ 
    \midrule
    Mean Cls. Acc. (ResNet34) & 61.3 & 68.0 & 68.4
    \\
    Overall Acc. (ResNet34) & 71.9 & 76.1 & 76.1
    \\
    Mean Cls. Acc. (ResNet50) & 58.1 & 65.0 & 70.4
    \\
    Overall Acc. (ResNet50) & 68.7 & 74.1 & 78.6
    \\
    Mean Cls. Acc. (ResNet101) & 68.0 & 66.1 & 69.6
    \\
    Overall Acc. (ResNet101) & 76.8 & 76.8 & 79.0
    \\
    \bottomrule
    \end{tabular}
\end{table}

\subsection{Discussion}

Besides the above species recognition results, we also discuss the effects of augmentation, and the supported research of our dataset and benchmark. 
The former is  for comparisons among none augmentation, a single type of augmentation (random flip) and two types of augmentations (random flip, Gaussian blur). The latter explains some crucial research which can be benefited from our work.


\textbf{Augmentation.}
As mentioned above, we augmented data during training. We implemented two types of augmentation which are random flip (vertical and horizontal) and Gaussian blur. We use the ResNet series in the case of pre-training initialization for this experiment. We observed from Table \ref{table:augdistricmp} that the augmentation is able to promote the classification accuracy. The complete augmentation (random flip, Gaussian blur) attains a better accuracy than both the single augmentation (random flip) and none augmentation. The single flip augmentation usually obtains a better accuracy than none augmentation. Though there is some improvement, it is not significant and the highest accuracy is still below $80.0\%$. This further demonstrates the challenges of our proposed dataset, and that augmentation is difficult to compensate the challenges.

\textbf{Research to support.} Our proposed dataset and benchmark can serve many relevant research activities, and some are listed as follows.
\begin{itemize}
    \item \textit{Pest control.} Crop rotation is the practice of planting different crops sequentially on the same plot of land, in order to improve soil health and control pest. The success of crop rotation relies on a proper selection of an alternative non-host crop for the pest. For the control of plant-parasitic nematodes, the nematode species as prior information is essential, since different species usually have differing host ranges.
    
    \item \textit{Soil ecology.} Soil nematodes are abundant in number and vary sensitively to pollutants and environmental disturbance. Therefore the presence and the abundance degree of certain soil nematode species is one of the most important bio-indicators to evaluate soil health, quality, and the physical or pollution-induced disturbances. In this direction, species recognition is critical and our work supports it.
    
    \item \textit{Bio-geography.} Many nematode species are cosmopolitan. Their worldwide distribution are affected by some biological and geographic factors. The understanding of the involved species and their assemblage (e.g. species recognition) is key to interpret the interactions of these factors, and how they contribute to the environment. Our work supports this direction as well.
    
    \item \textit{3D modeling.} 2D image based understanding of nematodes might lose some important information. 3D reconstruction and rendering of nematodes would provide more freedom for researchers' understanding of nematodes. These can be based on our proposed images dataset (e.g. 3D reconstruction from 2D images).
    
\end{itemize}

\subsection{Future Work}
The above results and discussion reveal the challenges of our dataset, especially for training a CNN from scratch. It will motivate researchers to analyze the properties of our data and propose innovative techniques (e.g. CNNs, data balance methods etc), in order to attain higher classification accuracy. 
In the future, we would like to add new data, and invite users to submit their test data, to our dataset. 
With more samples and species gathered, this dataset will be expanded to advance the species recognition of nematodes in the foreseeable future.

\section{Conclusion}
\label{sec:conclusion}

In this paper, we have presented an open-access imaging dataset for nematode recognition. The dataset has been collected and annotated manually, with paying rigorous efforts and time. We have investigated the efficacy of the state-of-the-art deep learning methods on nematode recognition, a \textit{biological computing} task. We have conducted a benchmark of adopting various state-of-the-art Convolutional Neural Networks (CNNs) on this dataset. We also discussed and analyzed the results. We found  augmentation and pre-training initialization can help boost the performance of species recognition. However, there is still great space to improve, given the current highest overall accuracy $79.0\%$. It is possible to improve the performance by designing sophisticated deep learning networks. We hope our dataset and benchmark will be instructive to relevant researchers from different fields in their future research.


\bibliographystyle{splncs04}
\bibliography{mybibliography}

\begin{thebibliography}{10}
\providecommand{\url}[1]{\texttt{#1}}
\providecommand{\urlprefix}{URL }
\providecommand{\doi}[1]{https://doi.org/#1}

\bibitem{pytorch}
Pytorch. \url{https://github.com/pytorch} (2020)

\bibitem{akintayo2018deep}
Akintayo, A., Tylka, G.L., Singh, A.K., Ganapathysubramanian, B., Singh, A.,
  Sarkar, S.: A deep learning framework to discern and count microscopic
  nematode eggs. Scientific reports  \textbf{8}(1),  1--11 (2018)

\bibitem{anderson2000nematode}
Anderson, R.C.: Nematode parasites of vertebrates: their development and
  transmission. Cabi (2000)

\bibitem{andrassy2005free}
Andr{\'a}ssy, I.: Free-living nematodes of Hungary:(Nematoda errantia). 1.
  Hungarian Natural History Museum (2005)

\bibitem{blaxter2004promise}
Blaxter, M.L.: The promise of a dna taxonomy. Philosophical Transactions of the
  Royal Society of London. Series B: Biological Sciences  \textbf{359}(1444),
  669--679 (2004)

\bibitem{bongers1988nematoden}
Bongers, T.: De nematoden van Nederland: een identificatietabel voor de in
  Nederland aangetroffen zoetwater-en bodembewonende nematoden. Koninklijke
  Nederlandse Natuurhistorische Vereniging Zeist, The Netherlands (1988)

\bibitem{canny1986computational}
Canny, J.: A computational approach to edge detection. IEEE Transactions on
  pattern analysis and machine intelligence (6),  679--698 (1986)

\bibitem{chen2020cnn}
Chen, L., Strauch, M., Daub, M., Jiang, X., Jansen, M., Luigs, H.G.,
  Schultz-Kuhlmann, S., Kr{\"u}ssel, S., Merhof, D.: A cnn framework based on
  line annotations for detecting nematodes in microscopic images. In: 2020 IEEE
  17th International Symposium on Biomedical Imaging (ISBI). pp. 508--512. IEEE
  (2020)

\bibitem{xception}
Chollet, F.: Xception: Deep learning with depthwise separable convolutions. In:
  Proceedings of the IEEE conference on computer vision and pattern
  recognition. pp. 1251--1258 (2017)

\bibitem{chou2017edge}
Chou, Y., Lee, D.J., Zhang, D.: Edge detection using convolutional neural
  networks for nematode development and adaptation analysis. In: International
  Conference on Computer Vision Systems. pp. 228--238. Springer (2017)

\bibitem{coomans2002present}
Coomans, A.: Present status and future of nematode systematics. Nematology
  \textbf{4}(5),  573--582 (2002)

\bibitem{imagenet_cvpr09}
Deng, J., Dong, W., Socher, R., Li, L.J., Li, K., Fei-Fei, L.: {ImageNet: A
  Large-Scale Hierarchical Image Database}. In: CVPR09 (2009)

\bibitem{derycke2008disentangling}
Derycke, S., Fonseca, G., Vierstraete, A., Vanfleteren, J., Vincx, M., Moens,
  T.: Disentangling taxonomy within the rhabditis (pellioditis) marina
  (nematoda, rhabditidae) species complex using molecular and morhological
  tools. Zoological journal of the Linnean Society  \textbf{152}(1),  1--15
  (2008)

\bibitem{erwin1982tropical}
Erwin, T.L.: Tropical forests: their richness in coleoptera and other arthropod
  species. The Coleopterists Bulletin  (1982)

\bibitem{floyd2002molecular}
Floyd, R., Abebe, E., Papert, A., Blaxter, M.: Molecular barcodes for soil
  nematode identification. Molecular ecology  \textbf{11}(4),  839--850 (2002)

\bibitem{resnet}
He, K., Zhang, X., Ren, S., Sun, J.: Deep residual learning for image
  recognition. In: Proceedings of the IEEE conference on computer vision and
  pattern recognition. pp. 770--778 (2016)

\bibitem{holladay2016high}
Holladay, B.H., Willett, D.S., Stelinski, L.L.: High throughput nematode
  counting with automated image processing. BioControl  \textbf{61}(2),
  177--183 (2016)

\bibitem{holterman2006phylum}
Holterman, M., van~der Wurff, A., van~den Elsen, S., van Megen, H., Bongers,
  T., Holovachov, O., Bakker, J., Helder, J.: Phylum-wide analysis of ssu rdna
  reveals deep phylogenetic relationships among nematodes and accelerated
  evolution toward crown clades. Molecular biology and evolution
  \textbf{23}(9),  1792--1800 (2006)

\bibitem{javer2018identification}
Javer, A., Brown, A.E., Kokkinos, I., Rittscher, J.: Identification of c.
  elegans strains using a fully convolutional neural network on behavioural
  dynamics. In: Proceedings of the European Conference on Computer Vision
  (ECCV). pp.~0--0 (2018)

\bibitem{alexnet}
Krizhevsky, A., Sutskever, I., Hinton, G.E.: Imagenet classification with deep
  convolutional neural networks. In: Proceedings of the 25th International
  Conference on Neural Information Processing Systems - Volume 1. p.
  1097–1105. NIPS'12, Curran Associates Inc., Red Hook, NY, USA (2012)

\bibitem{lambshead1993recent}
Lambshead, P.: Recent developments in marine benthic biodiversity reserch.
  Oceanis  \textbf{19},  5--24 (1993)

\bibitem{liu2017-2-projection}
{Liu}, M., {Wang}, X., {Liu}, X., {Zhang}, H.: Classification of multi-focal
  nematode image stacks using a projection based multilinear approach. In: 2017
  IEEE International Conference on Image Processing (ICIP). pp. 595--599 (Sep
  2017). \doi{10.1109/ICIP.2017.8296350}

\bibitem{Liu2017-1-imagefusion}
{Liu}, M., {Wang}, X., {Zhang}, H.: A multi-direction image fusion based
  approach for classification of multi-focal nematode image stacks. In: 2017
  IEEE International Conference on Image Processing (ICIP). pp. 3835--3839 (Sep
  2017). \doi{10.1109/ICIP.2017.8297000}

\bibitem{liu2010-xraymultilinear}
Liu, M., Roy-Chowdhury, A.K.: Multilinear feature extraction and classification
  of multi-focal images, with applications in nematode taxonomy. In: 2010 IEEE
  Computer Society Conference on Computer Vision and Pattern Recognition. pp.
  2823--2830. IEEE (2010)

\bibitem{liu2010-xray}
Liu, M., Roy-Chowdhury, A.K., Yoder, M., De~Ley, P.: Multi-focal nematode image
  classification using the 3d x-ray transform. In: 2010 IEEE International
  Conference on Image Processing. pp. 269--272. IEEE (2010)

\bibitem{liu2018-projection}
Liu, M., Wang, X., Liu, K., Liu, X.: Multi-focal nematode image stack
  classification using a projection-based multi-linear method. Machine Vision
  and Applications  \textbf{29}(1),  135--144 (2018)

\bibitem{liu2017-informationfusion}
Liu, M., Wang, X., Zhang, H.: Classification of nematode image stacks by an
  information fusion based multilinear approach. Pattern Recognition Letters
  \textbf{100},  22--28 (2017)

\bibitem{liu2018-cnn}
Liu, M., Wang, X., Zhang, H.: Taxonomy of multi-focal nematode image stacks by
  a cnn based image fusion approach. Computer methods and programs in
  biomedicine  \textbf{156},  209--215 (2018)

\bibitem{nadler2002species}
Nadler, S.: Species delimitation and nematode biodiversity: phylogenies rule.
  Nematology  \textbf{4}(5),  615--625 (2002)

\bibitem{nguyen2006improved}
Nguyen, H.T., Ji, Q.: Improved watershed segmentation using water diffusion and
  local shape priors. In: 2006 IEEE Computer Society Conference on Computer
  Vision and Pattern Recognition (CVPR'06). vol.~1, pp. 985--992. IEEE (2006)

\bibitem{nicol2011current}
Nicol, J., Turner, S., Coyne, D., Den~Nijs, L., Hockland, S., Maafi, Z.T.:
  Current nematode threats to world agriculture. In: Genomics and molecular
  genetics of plant-nematode interactions, pp. 21--43. Springer (2011)

\bibitem{ochoa2007contour}
Ochoa, D., Gautama, S., Vintimilla, B.: Contour energy features for recognition
  of biological specimens in population images. In: International Conference
  Image Analysis and Recognition. pp. 1061--1070. Springer (2007)

\bibitem{riddle1997developmental}
Riddle, D.L., Blumenthal, T., Meyer, B.J., Priess, J.R.: Developmental Genetics
  of the Germ Line--C. elegans II. Cold spring harbor laboratory press (1997)

\bibitem{Rizvandi-3}
{Rizvandi}, N.B., {Pizurica}, A., {Philips}, W., {Ochoa}, D.: Edge linking
  based method to detect and separate individual c. elegans worms in culture.
  In: 2008 Digital Image Computing: Techniques and Applications. pp. 65--70
  (Dec 2008). \doi{10.1109/DICTA.2008.87}

\bibitem{Rizvandi-2}
Rizvandi, N.B., Pi{\v{z}}urica, A., Philips, W.: Automatic individual detection
  and separation of multiple overlapped nematode worms using skeleton analysis.
  In: Campilho, A., Kamel, M. (eds.) Image Analysis and Recognition. pp.
  817--826. Springer Berlin Heidelberg, Berlin, Heidelberg (2008)

\bibitem{rizvandi2008machine}
Rizvandi, N.B., Pizurica, A., Philips, W.: Machine vision detection of isolated
  and overlapped nematode worms using skeleton analysis. In: 2008 15th IEEE
  International Conference on Image Processing. pp. 2972--2975. IEEE (2008)

\bibitem{seinhorst1962killing}
Seinhorst, J.: On the killing, fixation and transferring to glycerin of
  nematodes. Nematologica  \textbf{8}(1),  29--32 (1962)

\bibitem{silva2003intelligent}
Silva, C., Magalhaes, K., Neto, A.D.: An intelligent system for detection of
  nematodes in digital images. In: Proceedings of the International Joint
  Conference on Neural Networks, 2003. vol.~1, pp. 612--615. IEEE (2003)

\bibitem{vgg}
Simonyan, K., Zisserman, A.: Very deep convolutional networks for large-scale
  image recognition. In: Bengio, Y., LeCun, Y. (eds.) 3rd International
  Conference on Learning Representations, {ICLR} 2015, San Diego, CA, USA, May
  7-9, 2015, Conference Track Proceedings (2015),
  \url{http://arxiv.org/abs/1409.1556}

\bibitem{sohlenius1987vertical}
Sohlenius, B., Sandor, A.: Vertical distribution of nematodes in arable soil
  under grass (festuca pratensis) and barley (hordeum distichum). Biology and
  Fertility of Soils  \textbf{3}(1-2),  19--25 (1987)

\bibitem{inception}
Szegedy, C., Liu, W., Jia, Y., Sermanet, P., Reed, S., Anguelov, D., Erhan, D.,
  Vanhoucke, V., Rabinovich, A.: Going deeper with convolutions. In:
  Proceedings of the IEEE conference on computer vision and pattern
  recognition. pp.~1--9 (2015)

\bibitem{wang2020celeganser}
Wang, L., Kong, S., Pincus, Z., Fowlkes, C.: Celeganser: Automated analysis of
  nematode morphology and age. In: Proceedings of the IEEE/CVF Conference on
  Computer Vision and Pattern Recognition Workshops. pp. 968--969 (2020)

\bibitem{zhou2006automatic}
Zhou, B.T., Baek, J.H.: An automatic nematode identification method based on
  locomotion patterns. In: International Conference on Intelligent Computing.
  pp. 372--380. Springer (2006)

\end{thebibliography}

\end{document}